\documentclass[conference,10pt]{IEEEtran}
\IEEEoverridecommandlockouts
\usepackage{cite}
\usepackage{amsmath,amssymb}
\usepackage{graphicx}
\usepackage{booktabs}
\usepackage{multirow}
\usepackage{xcolor}
\usepackage{algorithm}
\usepackage{algpseudocode}
\usepackage{hyperref}

\newcommand{\sys}{RTLDetective}

\begin{document}
\raggedbottom

\title{\sys: Agentic RTL Debugging through Dataflow Analysis and Simulation-Guided Fault Localization}

\author{
\IEEEauthorblockN{Neusha Javidnia\textsuperscript{*}, Ruisi Zhang, Farinaz Koushanfar}
\IEEEauthorblockA{ECE Department \\
UC San Diego\\
}
\thanks{\textsuperscript{*}Corresponding author: Neusha Javidnia (njavidnia@ucsd.edu).}
}

\maketitle

\begin{abstract}
Large language model (LLM) agents have shown promise in RTL debugging, yet accurately localizing functional errors and generating reliable repairs remain challenging. We propose RTLDetective, a multi-agent framework that integrates signal dependency analysis, simulation-guided fault localization, and knowledge-based reasoning for automated RTL debugging and repair. RTLDetective constructs signal dependency graphs from RTL code and extracts failure information from waveform traces. A lightweight localizer agent predicts expected intermediate signal values and evaluates them through targeted simulation interventions, iteratively probing signal dependencies to isolate suspicious logic and identify potential fault locations. A debug agent combines this evidence with static analysis, a curated knowledge base of hardware design principles and RTL bug patterns, and reasoning to generate corrected RTL. To reduce inference overhead, the framework reuses cached context across debugging iterations. We evaluate RTLDetective on the MEIC benchmark, achieving an 86.1\% fix rate, 13.9 percentage points above the best baseline, while
making fewer LLM calls.
\end{abstract}

\section{Introduction}
\label{sec:introduction}

Large language models (LLMs) have advanced rapidly in recent
years and are increasingly used to assist hardware design,
from generating register-transfer level (RTL) code to
verification and debugging~\cite{chipchat, rtlcoder, chipnemo, assertllm, meic}.
RTL debugging is particularly important, as functional
verification accounts for a substantial portion of hardware
development effort, with engineers spending significant time
identifying and correcting design errors~\cite{wilson2024}.
Unlike syntax errors, which typically produce explicit compiler
diagnostics, functional errors can propagate through multiple
signals, registers, and module boundaries before becoming
observable at the outputs. Identifying their root causes therefore requires understanding both the design structure and its behavior over time.

Despite recent advances, applying LLMs to RTL debugging
faces two practical challenges. First, hardware description
languages (HDLs) are less represented in LLM training data
than general-purpose programming languages, limiting their
hardware-specific knowledge~\cite{llmfail}. Prior work has
addressed this limitation through continued pretraining, supervised fine-tuning,
and preference optimization~\cite{yao2025lik, veridebug, arsp, origen, hlsdebugger} for bug localization
and repair. However, these approaches require specialized
training data and additional computational resources,
making adaptation to new base models costly. Second, RTL
designs often contain proprietary intellectual property,
raising concerns about sharing design details with
cloud-hosted commercial models~\cite{chipnemo}. Together,
these challenges motivate debugging approaches that can
effectively leverage moderately sized, locally deployed
open-weight models without additional training.

Classical repair tools localize faults through program analysis, using data- and
control-flow dependencies~\cite{ahmad2022cirfix} or signal-value
transitions~\cite{strider}. Agentic frameworks instead integrate LLMs with compilers and
simulators in an iterative debugging loop, enabling them to propose repairs, evaluate
the results using tool feedback, and progressively refine their
solutions~\cite{rtlfixer, r3a}. MEIC~\cite{meic} and UVLLM~\cite{uvllm} iterate on
compiler and simulation results to guide repair, and R3A~\cite{r3a} pairs multi-agent
fault localization with tree-of-thoughts patch search. To point the LLM at the right
code, VerilogCoder~\cite{verilogcoder} traces signal dependencies back from a mismatched
output and supplies their waveforms, Veri-Sure~\cite{verisure} combines waveform
analysis with static dependency slicing, and VeriPilot~\cite{veripilot} compares
internal signals against a golden reference model to find where execution first
diverges.

Despite these advances, reliably identifying the root cause of functional failures remains challenging. Simulation reveals incorrect outputs but does not directly identify the faulty internal logic. Dependency tracing narrows the search, but without a golden reference, the expected values of intermediate signals remain unknown. Existing LLM-based approaches generally rely on reasoning over traces and dependencies rather than directly testing predicted signal values through simulation. Moreover, large RTL files and waveform quickly overwhelm the model's context and can distract the agent from the root cause, leading to incorrect localization and repeated repair attempts.

To address these challenges, we propose \sys{}, an agentic RTL debugging framework that combines hierarchical signal dependency analysis with simulation-guided fault localization. Given a failing testbench check, \sys{} builds a signal dependency graph across module boundaries and extracts the waveform transitions leading to the failure. A lightweight \textit{Localizer Agent} predicts the expected values of intermediate signals and tests them by forcing each predicted value in simulation to check whether the failure is resolved. Based on simulation feedback, the agent refines its predictions to narrow down the faulty logic. A larger \textit{Debug Agent} then uses the localized faults, design specification, and relevant debugging knowledge to generate RTL repairs, which are validated through simulation. Our main contributions are:

\begin{figure*}[t]
\centering
\includegraphics[width=0.9\textwidth]{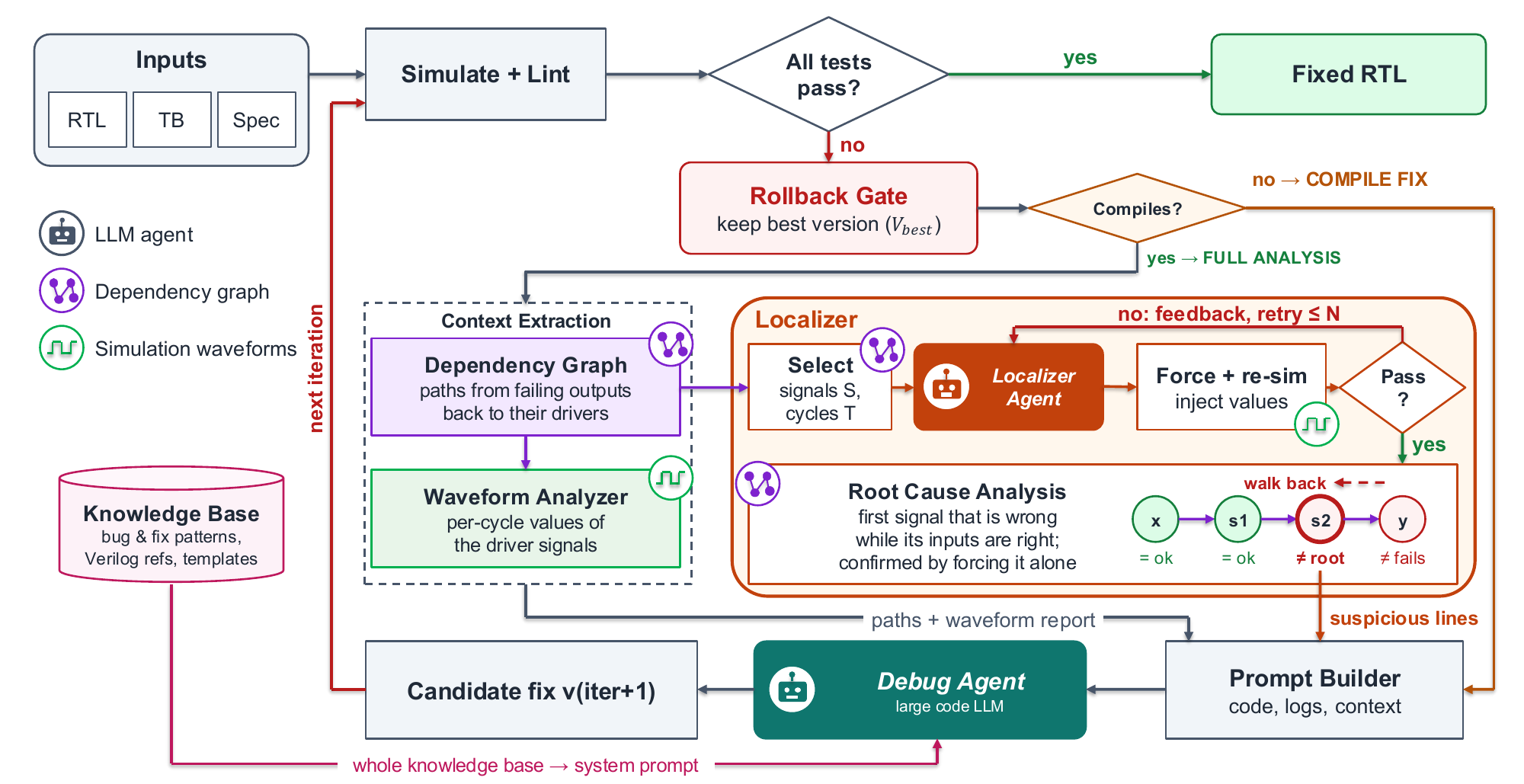}
\caption{Overview of \sys{}, a dual-agent RTL debugging framework
supported by a curated knowledge base. Each iteration evaluates
the current RTL candidate through compilation, static analysis,
and simulation. \emph{Context Extraction} constructs
a signal dependency graph, traces dependencies backward from
failing outputs, and summarizes waveforms leading up to the
first failure. The \emph{Localizer Agent} predicts expected
values of intermediate signals and tests them through
counterfactual simulation. \emph{Root Cause Analysis} uses
the validated predictions to identify suspicious logic.
The \emph{Debug Agent} uses the localized fault information
and retrieved RTL debugging knowledge to generate the next
repair candidate.}
\label{fig:overview}
\end{figure*}

\begin{itemize}
    \item A \textit{counterfactual fault localization} method that tests predicted internal signal values through targeted simulation to identify faulty logic without requiring a golden reference model.

    \item A multi-agent RTL debugging framework that integrates a lightweight \textit{Localizer Agent} with a \textit{Debug Agent}, using cross-module signal dependency analysis, compact waveform-based failure context, and prefix caching to support iterative repair.

    \item An evaluation on the MEIC benchmark using the same 30B repair model across all methods. \sys{} achieves an 86.1\% fix rate (FR), 13.9 percentage points above the best baseline. With three iterations, it outperforms the best baseline at ten iterations using less than half the tokens. On challenging designs, it fixes 83 of 106 bugs, compared with 59 for MEIC, while localization improves functional FR by 7.4 percentage points.

\end{itemize}

\section{Methodology}
\label{sec:method}

\subsection{\sys{} Overview}
\label{sec:overview}

Fig.~\ref{fig:overview} illustrates the overall workflow of
\sys{}, a dual-agent framework for automated RTL debugging.
The framework consists of two LLM agents: a
\emph{Localizer Agent}, which predicts expected internal
signal behavior, and a \emph{Debug Agent}, which
generates targeted RTL repairs. The \emph{Debug Agent} is further
supported by a curated knowledge base of RTL bug patterns
and design guidelines. Given a buggy RTL design, its
specification, and a testbench, \sys{} iteratively localizes
functional errors and generates repair candidates.

Each iteration begins with compilation, static analysis,
and simulation of the current RTL candidate.
The \emph{Failure Context Extraction} constructs a signal
dependency graph and extracts relevant waveform information
leading up to the failure. Using this context, the
\emph{Localizer Agent} predicts the expected values of intermediate
signals driving the failing outputs, which are evaluated
through counterfactual simulation. The \emph{Root Cause Analysis}
module uses the validated predictions to identify suspicious
logic. Next, the \emph{Debug Agent} combines the localized
fault information with retrieved debugging knowledge to
generate the next repair candidate. Finally, a rollback mechanism
preserves the best performing RTL version throughout the
iterations. This process repeats until all testbench checks
pass or the iteration budget is exhausted.

\subsection{Failure Context Extraction}
\label{sec:context}
For each candidate design, the failure context extractor collects three kinds of evidence: (1) compilation and simulation results, along with a rule-based static analyzer that identifies likely RTL bug patterns without LLM involvement; (2) a signal-dependency graph that traces failing outputs back to their driving signals across modules; and (3) a waveform analyzer that identifies where and when failures occur, condensing VCD traces into the most relevant signal values and transitions.

\subsubsection{Compilation, Simulation, and Static Analysis}
\label{sec:toolcontext}
Following prior agentic debugging frameworks, \sys{} grounds the agent in
tool feedback \cite{rtlfixer, mage}. Each candidate is compiled with Icarus Verilog \cite{iverilog}, and any
compilation errors and warnings are reported directly. If compilation
succeeds, the design is simulated against its testbench. The resulting log
records the overall pass rate and, for every failing test case, the expected
and observed output values.

In addition, \sys{} runs a static analyzer that applies regex-based and
structural rules to the RTL source. It detects common bug patterns without
simulation or synthesis. The rules flag:
(i) blocking assignments in sequential blocks;
(ii) non-blocking assignments in combinational blocks;
(iii) incomplete sensitivity lists;
(iv) \texttt{case} statements without a \texttt{default} branch;
(v) signals with multiple drivers;
(vi) bit-width mismatches between operands;
(vii) clock or reset signals in a sensitivity list without
\texttt{posedge}/\texttt{negedge}; and
(viii) combinational blocks with an \texttt{if} lacking an \texttt{else}
and no default assignment, which may infer a latch.
Each finding is reported with its source line, a severity level, and a
short explanation, and is passed to the \emph{Debug Agent} along with the
simulation results.

\subsubsection{Signal Dependency Graph Construction}
\label{sec:dependency_graph}
Starting from the simulation log, \sys{} identifies the
failing signals and constructs a hierarchical signal
dependency graph $G=(V,E)$ using \textit{Pyverilog} \cite{pyverilog}. Each node
represents an RTL signal, and a directed edge $u \to s$
indicates that $u$ can influence $s$. The graph captures
both data dependencies and control dependencies. Edges are classified
as sequential or combinational based on the corresponding
RTL assignments. Submodules are recursively elaborated,
with port connections linking signals across module
boundaries.

For each failing signal, \sys{} performs a depth-first
search to trace dependencies backward toward their
source signals, including feedback loops. The resulting
paths determine which signals are retained for waveform
analysis (Sec.~\ref{sec:waveform}) and provide the
structural context and candidate signals for the
\emph{Localizer Agent} during fault localization
(Sec.~\ref{sec:localize}).

\subsubsection{Waveform Analysis}
\label{sec:waveform}

The simulation log and static analysis show an output is wrong, but not how the error spread through the design's internal state. Building on prior waveform tracing~\cite{verilogcoder} in agentic frameworks, \sys{} further reduces redundant waveform information to provide the \emph{Debug Agent} with a compact view of the signals and cycles leading to functional failures.

The analyzer parses the VCD file generated during simulation and retains the DUT ports and internal signals along the dependency paths of failing outputs (Sec.~\ref{sec:dependency_graph}). For sequential designs, signals are sampled immediately before each rising clock edge, while combinational designs are sampled after input changes once values have settled. The analysis is limited to the interval from the last passing check (or reset release) to the first failing check. To compress the trace, constant and periodic signals are summarized, and rows without changes are removed. The resulting transition table provides the \emph{Debug Agent} with a compact view of signal behavior which directs it to where the bug first appears.

\subsection{Counterfactual Fault Localization}
\label{sec:localize}

Context extraction identifies the signals that influence a
failing output and their behavior over time, but does not
determine which signals are incorrect or what their expected
values should be. Since errors can propagate through signal
dependencies, an incorrect value may originate from faulty
logic or an earlier error. To distinguish between them,
\sys{} uses counterfactual simulation to test predicted
corrections to internal signals. The \emph{Localizer Agent}
first predicts the expected signal values based on the
specification. These are tested by temporarily forcing the values onto the corresponding internal signals during the selected cycles and releasing them afterward. Simulation feedback is then used to refine the predictions. Finally, the validated predictions are compared with the waveform to identify root cause candidates and
their RTL locations for the \emph{Debug Agent}. Alg.~\ref{alg:localize}
summarizes the procedure.

\algrenewcommand\algorithmicrequire{\textbf{Input:}}
\algrenewcommand\algorithmicensure{\textbf{Output:}}

\begin{algorithm}[t]
\caption{Counterfactual Fault Localization}
\label{alg:localize}
\small
\begin{algorithmic}[1]
\Require RTL $D$, testbench $TB$, specification $spec$,
graph $G$, waveform $W$, failing outputs $O$, max rounds $N$
\Ensure Localization report $L$, acceptance flag $accepted$
\State $S,T \gets \textsc{SelectSignalsAndCycles}(O,G,TB)$
\State $K \gets \textsc{InferKnownValues}(TB,W,G,S,T)$
\State $M \gets \textsc{BuildPrompt}(spec,D,G,S,T,K)$
\State $accepted \gets \textbf{false}$
\For{$r=1$ \textbf{to} $N$}
    \State $P \gets \textsc{LocalizerAgent}(M)$
    \State $res \gets \textsc{ForcedSimulate}(D,TB,P)$
    \If{$res.pass$}
        \State $accepted \gets \textbf{true}$; \textbf{break}
    \EndIf
    \State $M \gets M \cup \textsc{Feedback}(res,P)$
\EndFor
\State $C \gets \{(s,t) \in S \times T \mid P_s[t] \neq \bot,\ P_s[t] \neq W_s[t]\}$
\State $R \gets \textsc{FindRootCandidates}(C,G)$
\State $L \gets \emptyset$
\ForAll{$(s,t) \in R$}
    \State $confirmed \gets \textsc{ForcedSimulate}(D,TB,P|_s).pass$
    \State $L \gets L \cup \{(s,\,t,\,\textsc{DriverLines}(s),\,confirmed)\}$
\EndFor
\State \Return $L,\ accepted$
\end{algorithmic}
\end{algorithm}

\subsubsection{Intermediate Signal Prediction}
\label{sec:predict}

Starting from the failing outputs $O$, \sys{} selects
a set of driving signals $S$ through breadth-first
backward traversal of the dependency graph $G$.
Signals are selected across different dependency paths
to avoid focusing on a single branch. The cycle set $T$
covers the cycles leading up to the first failing check,
including the last passing check as a reference.

Before invoking the LLM, \sys{} fills in values that
can be inferred from the testbench and waveform,
including expected output values, observed values at
the last passing check, and values derived from direct
signal connections. The \emph{Localizer Agent} receives
these known values, the specification, RTL code,
signal dependencies, input stimulus, and a partially
filled signal-value table. It predicts the remaining
values based on the intended design behavior.
Values that cannot be determined are marked as unknown
($\bot$). The resulting $P_s[t]$ represents the
predicted value of signal $s$ at cycle $t$.

\subsubsection{Simulation-Based Hypothesis Validation}
\label{sec:validate}

The predicted values $P$ form a hypothesis about the expected internal behavior of the design. To test it, \sys{} re-runs the testbench, temporarily overriding the corresponding internal signals using \texttt{force} and \texttt{release} statements. The hypothesis is accepted if the original failing check passes without causing any earlier checks to fail. Otherwise, the simulator provides feedback on the remaining failing outputs and their expected and observed values. The \emph{Localizer Agent} uses this feedback to refine its predictions for up to $N$ rounds. If no hypothesis is accepted after $N$ rounds, the final prediction is still used for root cause analysis, but its results are marked as unaccepted.

\subsubsection{Root Cause Analysis}
\label{sec:rootcause}
\sys{} compares the predictions $P$ with the observed waveform $W$ to find
where the error first appears. A pair $(s,t)$ is a root cause candidate if
$P_s[t] \neq W_s[t]$ while none of its selected drivers mismatches at
cycle $t$, or at $t-1$ if $s$ is a register. Each candidate is then
re-simulated with its predictions $P|_s$ (the row of $P$ for signal $s$) forced, and is confirmed if this resolves the failing check
without breaking earlier checks. Finally, each candidate is mapped to the
RTL lines that drive it and passed to the \emph{Debug Agent} with its
observed and predicted values and confirmation status.

\subsection{Knowledge-Guided RTL Repair}
\label{sec:repair}
\subsubsection{Curated RTL Debugging Knowledge Base}
\label{sec:knowledgebase}
The \emph{Debug Agent} is supported by a curated knowledge base of RTL debugging expertise. It covers common syntax and functional bugs, including categories from MEIC \cite{meic}, their likely causes, and repair patterns with before-and-after examples. It also provides a structured debugging procedure, reference implementations for common RTL modules, and Verilog coding and synthesis guidelines. The knowledge base is compact (approximately 29\,KB) and is included directly in the agent's system prompt, making it available throughout the repair process without additional retrieval. The agent uses this knowledge to guide fault diagnosis and repair while treating the design specification as the primary reference for correct behavior.

\subsubsection{Repair Generation}
\label{sec:repairgen}
In each iteration, the \emph{Debug Agent} receives the design specification, buggy RTL code, static analysis results, and compilation and testbench feedback. For functional failures, it also receives the compressed waveform analysis (Sec.~\ref{sec:context}) and fault localization report (Sec.~\ref{sec:localize}). If compilation fails, the agent focuses on resolving compilation errors while preserving the intended functionality. Otherwise, it produces a structured repair report identifying each suspected bug, its RTL location and category, expected behavior, root cause, and proposed correction with supporting reasoning. The agent then generates the complete repaired RTL. By requiring an explanation of the intended behavior and root cause before proposing a fix, \sys{} encourages guided reasoning rather than corrections alone. The repaired design is then evaluated through compilation and simulation, and the process repeats until all testbench checks pass or the iteration limit is reached.

\textbf{Efficient Context Reuse.} Rather than treating each repair attempt independently, \sys{} maintains the agent's conversation history across iterations, allowing it to build on previous hypotheses, failed repairs, and simulation feedback without repeating earlier mistakes. To support this growing context efficiently, \sys{} leverages Key Value cache prefix reuse, caching the attention states of the shared prompt prefix, including the debugging knowledge base and prior interactions. As the conversation grows, previously processed context can be reused across successive requests \cite{vllm_prefix}, reducing redundant computation while preserving the agent's accumulated debugging knowledge.

\subsubsection{Rollback}
\label{sec:rollback}
LLM-generated repairs may introduce compilation errors or cause previously passing tests to fail. \sys{} therefore scores each candidate by evaluating it based on compilation success and testbench pass rate. It maintains the best-performing RTL version, $V_{\mathrm{best}}$. A candidate replaces $V_{\mathrm{best}}$ if its score is equal to or higher than the current best; otherwise, the change is reverted. Each iteration starts from $V_{\mathrm{best}}$, and the \emph{Debug Agent} receives feedback on rejected repairs and their test results to guide subsequent corrections and avoid repeating unsuccessful changes.

\section{Experiments}
\label{sec:experiments}
This section presents the benchmark, evaluation metrics,
baselines, and experimental setup, followed by an evaluation of debugging performance, computational efficiency, and ablation studies.

\begin{table*}[t]
\centering
\caption{RTL debugging performance on the MEIC benchmark.
All methods use Qwen3-Coder-30B-A3B with identical sampling settings.
FR denotes fix rate.}
\label{tab:main}
\setlength{\tabcolsep}{4pt}
\begin{tabular}{llcccccccc}
\toprule
& & \multicolumn{2}{c}{Syntax}
  & \multicolumn{2}{c}{Functional}
  & \multicolumn{2}{c}{Overall}
  & \multicolumn{2}{c}{Avg. LLM Calls/Bug*} \\
\cmidrule(lr){3-4}
\cmidrule(lr){5-6}
\cmidrule(lr){7-8}
\cmidrule(lr){9-10}
Method & Budget
& Fixed/Total & FR (\%)
& Fixed/Total & FR (\%)
& Fixed/Total & FR (\%)
& Repair & Auxiliary \\
\midrule
Qwen, no agent
    & 1 call
    & 50/113 & 44.2
    & 29/81 & 35.8
    & 79/194 & 40.7
    & 1.00 & 0 \\

UVLLM~\cite{uvllm} 
    & 4+4 iter. 
    & 70 / 113 & 61.9 
    & 44 / 81 & 54.3 
    & 114 / 194 & 58.8 
    & 5.78 
    & 0 \\

MEIC~\cite{meic}
    & 3 iter.
    & 84/113 & 74.3
    & 51/81 & 63.0
    & 135/194 & 69.6
    & 1.47 & 1.47 \\

MEIC~\cite{meic}
    & 10 iter.
    & 84/113 & 74.3
    & 56/81 & 69.1
    & 140/194 & 72.2
    & 3.60 & 3.60 \\

\midrule
\sys{}
    & 3 iter.
    & 99/113 & 87.6
    & 58/81 & 71.6
    & 157/194 & 80.9
    & \textbf{1.40} & 1.59 \\

\sys{}
    & 10 iter.
    & \textbf{106/113} & \textbf{93.8}
    & \textbf{61/81} & \textbf{74.1}
    & \textbf{167/194} & \textbf{86.1}
    & 2.07 & 3.12 \\
\bottomrule
\end{tabular}

\vspace{2pt}
\begin{minipage}{0.95\linewidth}
\footnotesize
\textit{*Note:} Repair denotes calls to the primary 30B model.
Auxiliary calls include additional pipeline calls in UVLLM,
scoring agent in MEIC, and localizer agent in \sys{}.
\end{minipage}
\end{table*}
\subsection{Experimental Setup}
\label{sec:setup}

\textbf{Benchmark.} We evaluate \sys{} on an extended version of the MEIC benchmark~\cite{meic}, comprising 194 buggy instances across 15 RTLLM designs~\cite{rtllm}. The benchmark contains 113 syntax errors and 81 functional bugs, with natural-language specifications and self-checking testbenches.

\textbf{Baselines.}
We compare \sys{} against three baselines:
\begin{itemize}
    \item {No Agent:} A single-pass LLM baseline that
    generates a repair from the design specification and buggy
    RTL, without simulation feedback, iterative refinement, or
    external tools. 

    \item {MEIC}~\cite{meic}: An iterative RTL repair
    framework consisting of a debug agent that proposes fixes
    using compiler and simulation feedback, and a scoring agent
    that evaluates and selects repair candidates.

    \item {UVLLM}~\cite{uvllm}: A multi-stage RTL repair
    framework that performs lint-guided syntax correction,
    identifies output mismatches against a reference (oracle)
    simulation, and localizes suspicious RTL statements for
    further repair.
\end{itemize}

\noindent\textbf{Models and Agent Settings.} All methods use \texttt{Qwen3-Coder
-30B-A3B-Instruct}\cite{qwen3coder}, a mixture-of-experts model with 30.5B parameters (3.3B active), served by vLLM~\cite{vllm} with prefix caching on two NVIDIA RTX A6000 GPUs. Generation uses temperature 0.2, top-$p$ 0.9, repetition penalty 1.1, and up to 16{,}384 output tokens. \sys{} additionally uses \texttt{Qwen3-8B}~\cite{qwen3} as the \emph{Localizer Agent}, deployed on a third A6000 with temperature 0.7, top-$p$ 0.95, top-$k$ 20, a maximum of 1{,}024 output tokens. Each localization attempt allows up to four prediction rounds. All methods use Icarus Verilog 13.0~\cite{iverilog} for compilation, simulation, and final evaluation.

\sys{} and MEIC are both evaluated with repair budgets of 3 and 10 iterations per bug. UVLLM uses its default budget of up to four output-mismatch repair iterations followed by up to four suspicious-line repair iterations, while the No Agent baseline makes a single attempt.

\noindent\textbf{Evaluation Metric.} We evaluate debugging effectiveness using Fix Rate (FR) ~\cite{meic, debugbench}, defined as the percentage of buggy RTL designs
whose final repaired versions pass all testbench cases:

\begin{equation}
    \mathrm{FR} = \frac{N_{\mathrm{fixed}}}{N_{\mathrm{total}}}
    \times 100\%,
\end{equation}
where $N_{\mathrm{fixed}}$ is the number of designs whose final
repaired RTL passes all testbench cases, and $N_{\mathrm{total}}$ is
the total number of buggy designs.

To assess computational efficiency, we report the average number of LLM calls, input and output tokens, and wall-clock time per bug. We separate repair calls to the main model from auxiliary calls, including MEIC's scoring agent, UVLLM's additional pipeline calls, and \sys{}'s \textit{Localizer Agent}. Wall-clock time includes model inference, compilation, simulation, and tool execution. To ensure reliable timing, all methods process one bug at a time on an idle server. Separately, we measure the latency benefit of prefix caching using time to first token (TTFT), comparing recorded \textit{Debug Agent} calls replayed with caching enabled and disabled.

\subsection{Main Results}
\label{sec:results}

\begin{table}[t]
\centering
\caption{Fixed bugs per design. $n$: buggy instances per design.}
\label{tab:per_design}
\setlength{\tabcolsep}{2.5pt}
\begin{tabular}{lrrrrrrr}
\toprule
 & & & & \multicolumn{2}{c}{MEIC} & \multicolumn{2}{c}{Ours} \\
\cmidrule(lr){5-6}\cmidrule(lr){7-8}
Design & $n$ & no-agent & UVLLM & 3 it. & 10 it. & 3 it. & 10 it. \\
\midrule
Accumulator           & 12 & 0  & 6  & 4  & 4  & 9  & \textbf{10} \\
8-bit adder           & 12 & \textbf{12} & 10 & 11 & \textbf{12} & \textbf{12} & \textbf{12} \\
32-bit adder          & 14 & 9  & 3  & 8  & 9  & \textbf{12} & \textbf{12} \\
Pipelined adder       & 12 & 11 & 9  & \textbf{12} & \textbf{12} & 11 & \textbf{12} \\
Arithmetic logic unit & 12 & 0  & 11 & \textbf{12} & \textbf{12} & \textbf{12} & \textbf{12} \\
Booth multiplier      & 13 & 9  & 10 & 12 & 12 & 12 & \textbf{13} \\
Pipelined multiplier  & 12 & 3  & 7  & 7  & 9  & 9  & \textbf{10} \\
16-bit divider        & 11 & 0  & 4  & 4  & 5  & 7  & \textbf{8}  \\
Radix-2 divider       & 15 & 0  & 2  & 8  & 8  & 11 & \textbf{12} \\
Parallel-to-serial    & 12 & 11 & 8  & \textbf{12} & \textbf{12} & \textbf{12} & \textbf{12} \\
Serial-to-parallel    & 13 & 0  & 7  & 7  & 7  & \textbf{9}  & \textbf{9}  \\
Width converter       & 11 & \textbf{10} & 9  & \textbf{10} & \textbf{10} & 8  & \textbf{10} \\
Frequency divider     & 14 & 13 & 12 & \textbf{14} & \textbf{14} & 13 & 13 \\
Traffic-light control & 15 & 1  & 10 & 10 & 10 & 9  & \textbf{11} \\
Asynchronous FIFO     & 16 & 0  & 6  & 4  & 4  & \textbf{11} & \textbf{11} \\
\midrule
\textbf{Total}        & 194 & 79 & 114 & 135 & 140 & 157 & \textbf{167} \\
\bottomrule
\end{tabular}
\end{table}

\begin{table*}[t]
\centering
\caption{Repair cost and efficiency on the MEIC benchmark,
averaged over all bugs.}
\label{tab:cost}
\setlength{\tabcolsep}{6pt}
\begin{tabular}{llccc}
\toprule
Method & Budget & Fixed (/194) $\uparrow$
& Tokens/Bug (k) $\downarrow$
& Time/Bug (s) $\downarrow$ \\
\midrule
Qwen, no agent     & 1 call        & 79  & 1.9  & 6.1 \\
UVLLM~\cite{uvllm} & 4+4 iter. & 114 & 8.8  & 24.2 \\
MEIC~\cite{meic}   & 3 iter.       & 135 & 10.6 & 10.0 \\
MEIC~\cite{meic}   & 10 iter.      & 140 & 64.6 & 28.4 \\
\midrule
\sys{}             & 3 iter.       & 157 & 31.2 {\small(6.3 loc.)}  & 40.0 \\
\sys{}             & 10 iter.      & \textbf{167} & 66.2 {\small(12.3 loc.)} & 71.4 \\
\bottomrule
\end{tabular}

\vspace{2pt}
\begin{minipage}{0.95\linewidth}
\footnotesize
\textit{Note:} Token counts include input and output tokens.
For \sys{}, values in parentheses indicate tokens used by
the localizer, already included in the total.
Wall-clock time includes model inference, compilation,
simulation, and tool execution.
\end{minipage}
\end{table*}

\noindent\textbf{Overall Repair Performance.} Table~\ref{tab:main} compares the fix rate (FR) of \sys{} against the baselines across syntax and functional bugs. With three iterations, \sys{} fixes 80.9\% of all bugs, outperforming MEIC by 11.3 percentage points at the same budget and by 8.7 points even
when MEIC uses ten iterations. At the full budget, \sys{} reaches 86.1\% FR, 13.9 points above the best baseline. The largest gain is on syntax bugs, where \sys{} reaches 93.8\% FR, 19.5 points above the best baseline. For functional bugs, \sys{} with three iterations already surpasses MEIC with ten (71.6\% vs.\ 69.1\%) and ultimately reaches 75.3\% FR.

To isolate the contribution of the \textit{Localizer Agent}, we compare separate
three-iteration runs of \sys{} with and without localization. Without it, the functional FR drops from 66.7\% to 59.3\%, while syntax bug performance remains unchanged. Since the localizer is invoked only for functional bugs, its
benefit lies in identifying their root causes through simulation feedback.

\noindent\textbf{Per Design.}
Table~\ref{tab:per_design} reports per-design results. \sys{} achieves the highest number of successful repairs on 14 of the 15 designs, including 9 where it outperforms all baselines. Notably, its advantage is mostly evident in challenging designs where \emph{no-agent} LLM repair performs poorly. On the eight designs where the \emph{no-agent} baseline fixes at most three bugs each, \sys{} repairs 83 of 106 instances, compared with 59 for MEIC with ten iterations and 53 for UVLLM. On the remaining designs, where direct repair is already effective, \sys{} and MEIC achieve comparable results. These findings highlight the benefit of \sys{} on designs that are difficult to repair through \emph{no-agent} reasoning alone.

\subsection{Cost and Efficiency}
Table~\ref{tab:main} reports LLM calls, while Table~\ref{tab:cost} summarizes token usage and runtime. With three iterations, \sys{} makes an average of 2.99 LLM calls per bug (1.40 repair and 1.59 localization calls) and uses 31.2k tokens, including 6.3k for localization. Compared with MEIC at ten iterations, \sys{} uses less than half the tokens (31.2k vs.\ 64.6k) while fixing more bugs. MEIC at the same budget and UVLLM use fewer tokens but also fix fewer bugs. Increasing \sys{}'s budget to ten iterations yields 10 additional fixes but more than doubles token usage to 66.2k per bug. This performance improvement comes with higher runtime: at the full budget, \sys{} takes 71.4\,s per bug, compared with 28.4\,s for MEIC.

Table~\ref{tab:kvcache} shows the effect of prefix caching on the
\textit{Debug Agent}. On average, 82\% of each prompt’s tokens are reused from the prefix cache. Caching reduces the mean time to first token (TTFT) by 3.9$\times$, from 2370\,ms to 612\,ms, saving 1.76\,s per call. The speed-up grows with each iteration, from 3.2$\times$ in the first to 4.6$\times$ in the third. This is because each iteration extends the prompt of the previous one and only the new part has to be processed. Since most of a call's time is spent decoding, which caching does not shorten, whole calls are 9.1\% faster (84.6 vs.\ 93.1 minutes over all calls).

\begin{table}[t]
\centering
\caption{Effect of prefix (KV) caching on time to first token (TTFT) of the
\textit{Debug Agent}.}
\label{tab:kvcache}
\begin{tabular}{lrrrrrr}
\toprule
 & & Prompt & Cached & \multicolumn{2}{c}{TTFT (ms)} & \\
\cmidrule(lr){5-6}
 & $n$ & (k tok) & (\%) & Off & On & Speed-up \\
\midrule
Iteration 1 & 155 & 13.2 & 78 & 1674 & 517 & 3.2$\times$ \\
Iteration 2 &  78 & 18.2 & 85 & 2574 & 619 & 4.2$\times$ \\
Iteration 3 &  57 & 24.6 & 88 & 3985 & 863 & 4.6$\times$ \\
\midrule
\textbf{All} & 290 & 16.8 & 82 & 2370 & 612 & \textbf{3.9$\times$} \\
\bottomrule
\end{tabular}
\end{table}

\subsection{Ablation Study}
\label{sec:ablation}

\begin{figure}[t]
  \centering
  \includegraphics[width=0.8\columnwidth]{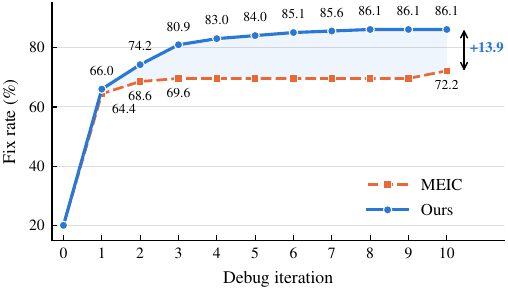}
  \caption{Fix rate (FR) versus debug iteration on the MEIC benchmark. \sys{}
  framework keeps improving until iteration~8 and ends 13.9 points higher.}
  \label{fig:fr_vs_iter}
\end{figure}

Fig.~\ref{fig:fr_vs_iter} shows the cumulative fix rate (FR) across debugging iterations. Both methods repair nearly two-thirds of the bugs in the first iteration (66.0\% vs.\ 64.4\%), primarily syntax errors. However, their performance diverges with additional iterations. MEIC plateaus at 69.6\% from iterations 3 to 9, whereas \sys{} continues to improve through iteration 8, using \textit{Counterfactual Fault Localization} and \textit{Failure Context Extraction} to guide functional bug repairs using simulation feedback. Notably, \sys{} achieves 80.9\% FR in just three iterations, surpassing its baseline. With the full budget, \sys{} reaches 86.1\% FR, outperforming MEIC by 13.9\%.

\section{Conclusion}
\label{sec:conclusion}

We presented \sys{}, a multi-agent framework for automated RTL debugging that validates fault hypotheses through targeted simulation. A lightweight \textit{Localizer Agent} predicts intermediate signal values along cross-module dependency paths and tests them through forced simulation, without requiring a golden reference model. A \textit{Debug Agent} uses the localized faults, compact failure context, and RTL debugging knowledge to generate targeted repairs. Evaluated on 194 bugs from the MEIC benchmark using the same open-weight 30B repair model across all methods, \sys{} achieves an 86.1\% fix rate, outperforming the best baseline by 13.9 percentage points. With only three iterations, it also surpasses the best baseline's ten-iteration result while using less than half the tokens. Future work will explore applying this model-agnostic fault validation approach to other LLM-based verification tasks, including assertion generation and testbench debugging.

\bibliographystyle{IEEEtran}
\bibliography{references}

\end{document}